# Experimental realization of on-chip topological nanoelectromechanical metamaterials

**Authors:** Jinwoong Cha[1,2], Kun Woo Kim[3], Chiara Daraio[2*]

**Affiliations:**

[1]Department of Mechanical and Process Engineering, ETH Zurich, Switzerland

[2]Engineering and Applied Science, California Institute of Technology, Pasadena, CA, USA

[3]Korea Institute for Advanced Study, Seoul, Republic of Korea

*Correspondence to: Prof. Chiara Daraio (daraio@caltech.edu)

**Topological mechanical metamaterials translate condensed matter phenomena, like non-reciprocity and robustness to defects, into classical platforms[1,2]. At small scales, topological nanoelectromechanical metamaterials (NEMM) can enable the realization of on-chip acoustic components, like unidirectional waveguides and compact delay-lines for mobile devices. Here, we report the experimental realization of NEMM phononic topological insulators, consisting of two-dimensional arrays of free-standing silicon nitride (SiN) nanomembranes that operate at high frequencies (10-20 MHz). We experimentally demonstrate the presence of edge states, by characterizing their localization and Dirac cone-like frequency dispersion. Our topological waveguides also exhibit robustness to waveguide distortions and**

**pseudospin-dependent transport. The suggested devices open wide opportunities to develop functional acoustic systems for high-frequency signal processing applications.**

Wave-guiding through a stable physical channel is strongly desired for reliable information transport in on-chip devices. However, energy transport in high-frequency mechanical systems, for example based on microscale phononic devices, is particularly sensitive to defects and sharp turns because of back-scattering and losses[3]. Two-dimensional topological insulators, first described as quantum spin hall insulators (QSHE) in condensed matter[4-7], demonstrated robustness and spin-dependent energy transport along materials' boundaries and interfaces. Translating these properties in the classical domain offers opportunity for scaling the size of acoustic components to on-chip device levels.

Topological phenomena have been shown in various architectured materials like photonic[8-11], acoustic[12,13], and mechanical[1,2,14-18] metamaterials. Especially, photonic systems have recently demonstrated the use of topological effects for lasing[19-21] and quantum interfaces[22]. However, acoustic and mechanical topological systems have so far been realized only in large-scale systems, like arrays of pendula[1], gyroscopic lattices[2], and arrays of steel rods[12], laser-cut plates[18] which require external driving systems. To fulfil their potential in device applications, mechanical topological systems need to be scaled to on-chip level for high-frequency transport.

Nanoelectromechanical systems[23,24] (NEMS) can be employed to build on-chip topological acoustic devices, thanks to their ability to transduce electrical signal into mechanical motion, which is essential in applications. NEMS systems with few degrees of freedom have already demonstrated quantum-analogous phenomena, like cooling and amplification[25], and Rabi-oscillation[26,27]. One-dimensional (1-D) nanoelectromechanical lattices (NEML) are a different class of NEMS devices used to study lattice dynamics for example, in waveguiding[28,29], and energy focusing[30]. Recently, 1-D NEMLs made of SiN nanomembranes have demonstrated active manipulation of phononic dispersion, leveraging electrostatic softening effects and nonlinearity[31].

Here, we fabricate and test high-frequency phononic topological insulators in engineered, two-dimensional, nano-electromechanical metamaterials. We employ an extended honeycomb lattice, which contains six lattice sites in a unit cell, satisfying $C_6$ symmetry (Supplementary information). This lattice exploits Brillouin zone-folding to demonstrate a double-Dirac cone at Γ point of the Brillouin zone (BZ). The zone-folding method has been recently used in various topological photonic[9,10], acoustic[12,13] and elastic[16] systems, by introducing the concept of pseudospins that satisfy Kramers theorem[7]. BZ folding allows to realize a pseudo-time reversal symmetry invariant system, where a time-reversal operator is defined from the symmetry ($C_6$) of the lattice. In our systems, the topological phase is controlled by the coupling strength, $t$, between unit cells within the extended honeycomb supercell and between adjacent cells, $t'$ (Fig. S1 and S2).

According to a classification method for topological phonons[15], extended honeycomb lattices are a class A*II* topological insulators, characterized by a $Z_2$ topological invariant with $U_T U_T^\dagger = -1$. Here, $U_T$ is the time-reversal operator, which is anti-unitary (Supplementary information). To calculate the topological invariant, we consider spin-Chern numbers from BHZ model[6] near the Γ point, as pseudospins are not fully preserved in entire BZ. The spin-Chern numbers $C_\pm = 0$ for $t > t'$, $C_\pm = \pm 1$ for $t < t'$ confirms the pseudospin-dependent edge states (Supplementary information). The $Z_2$ topological invariants are $v = (C_+ - C_-)/2$ (mod 2) = 0 for $t > t'$, and $v = 1$ for $t < t'$, and support different topological phases. The difference in the couplings, $t$ and $t'$, leads to a band gap opening at Γ point, with a dipole-to-quadrupole vibrational band inversion for a non-trivial lattice ($t > t'$) (Supplementary information). If a domain wall is formed between the two topologically distinct phases, gapless topological edge states emerge.

We realize these topological properties in our NEMM by periodically arranging etch holes, with 500 nm diameter, in an extended honeycomb lattice. The etch holes enable a buffered oxide etchant (BOE) to partially remove the sacrificial thermal oxide layer and release the SiN suspended membranes (Fig. 1a). We engineer the topological phases of the lattice, by changing the distance between etch holes, $w$. This strategy allows us to control the lattice couplings, $t$ and $t'$, from the overlaps between the circular etching paths from the neighboring etch holes (Extended Data Fig. 1). Recent reports[28-31] on 1-D NEML have exploited the isotropic nature of HF etching for device fabrication. Our NEMM consists of a periodic array of free-standing SiN nanomembrane forming a flexural phononic crystal. The average thickness of the nano-membranes is ~79 nm. The average vacuum gap distance between the SiN layer and the highly doped silicon substrate is ~147 nm. These values are estimated considering the partial etching rate of the SiN in the BOE etchant (~0.3 nm/min).

We perform finite element simulations using COMSOL©, to numerically compute frequency dispersion curves for a unit cell with a lattice parameter, $a = 18$ μm (Extended Data Fig. 2). We vary the distance between two neighboring holes, $w$, from 5.5 μm to 6.5 μm (Fig. 1b to 1e). For a unit cell with $w = 6.0$ μm $= a / 3$, a double Dirac-cone is present around 14.55 MHz at the Γ point of the BZ. (Fig. 1c). The frequency dispersion curves typically start from around 12 MHz, because of the presence of clamped boundaries. The frequency dispersion curves for $w = 5.5$ μm and 6.5 μm show the emergence of ~1.8 MHz-wide band gaps at the Γ point, ranging from 14 MHz to 15.8 MHz. The lattice with $w = 5.5$ μm exhibits two additional band gaps below and above the center band gap around 15 MHz (Fig. 1b), while the lattice with $w = 6.5$ μm (Fig. 1d) does not. The four vibrational modes, $p_x$, $p_y$, $d_{xy}$, and $d_{x^2-y^2}$, at the Γ point are degenerate at the Dirac point for the lattice with $w$ = 6 μm (Fig. 1c and 1e). The four degenerate modes are split into two separate degenerate modes, for $w$ < 6 μm and $w$ > 6 μm (Fig. 1e), opening a band gap. The band inversion between the dipole vibrational modes ($p_x$, $p_y$) and the quadrupole ones ($d_{xy}$, $d_{x^2-y^2}$) appears at the Γ point for $w$ > 6 μm, which supports the topological non-triviality of the lattice.

To study the topological properties of our NEMMs, we first fabricate a straight topological edge waveguide (Fig. 2a and 2b), formed at the interface of the topologically trivial ($w$ = 5.5 μm, Fig. 2c) and non-trivial ($w$ = 6.5 μm, Fig. 2d) lattices. Topological edge states do not exist at free-boundaries of our systems, owing to the lack of $C_6$ symmetry. The number of unit cells of each phase is approximately 200, so that the edge waveguide has 20 supercells with 18 μm one-dimensional lattice spacing. To characterize the edge states, we excite the flexural motion of the membranes by applying a dynamic electrostatic force, $F \propto (V_{DC} + V_{AC})^2$, to the excitation electrode. Here, $V_{DC}$ and $V_{AC}$ are DC and AC voltages, which are simulteneously applied between the excitation electrode and the grounded substrate (Fig. 2a). We perform measurements using a home-built Michelson interferometer with a balanced homodyne detection scheme (Methods). To obtain the dispersion curves of the edge states, we measure the frequency responses of 20 sites along the edge waveguide, by spatially scanning the measurement points (yellow strip, Fig. 2a) with a 18 μm step size (Extended Data Fig. 3). The Dirac-like edge state frequency disperison curves, isolated from the bulk dispersion, are present in the frequency range between 14.1 MHz and 15.8 MHz, showing a good agreement with the numerical dispersion curves (Fig. 2f). We also observe a defect mode at the crossing point of the edge state dispersion curves (Fig. 2e). This stems from a point defect mode from the boundary near the excitation region. The broken $C_6$ symmetry at the interfaces introduces a small band gap in the middle of the edge state dispersions (Fig. 2b). Despite the presence of the band gap, the defect mode is allowed to transmit non-negligible energy to the end of the waveguide owing to the long decay length of the evanescent mode.

We also characterize the localization of the edge states, by scanning the measurement point across the waveguide (yellow dashed-line AB in Fig. 2a) also with an 18 μm step size. The edge states are strongly localized (Fig. 2h) within ± 36 μm distance from the interface (Fig. 2g). Beyond this range, the frequency responses (Fig. 2g) start to show clear band gaps with similar positions and magnitude to the numerical dispersions, shown in Fig1b and 1d. The trivial lattice side presents three band gaps (Fig. 2g, left) and the non-trivial lattice side shows only one topological band gap (Fig. 2g, right), as predicted in the numerical frequency dispersion (Fig. 1b to 1d). The frequency responses show evidence of different topological phases in the two lattices, $w = 5.5$ and $6.5$ μm, confirming that the waveguiding effect is topological.

One remarkable feature of topological edge modes is their robustness to waveguide imperfections, like sharp corners. To study this, we additionaly fabricate a zig-zag waveguide with two sharp corners, with 60° angles (Fig. 3a). We perform steady-state and transient response measurements, by scanning the laser spot along the waveguide with 18 μm scanning steps. The frequency dispersion from the steady state measurements confirms the presence of the topological edge states (Extended Data Fig. 4). To validate the back-scattering immunity, we perform transient response measurements with propagating pulses, because the steady-state responses include signals from boundary-scattered waves. For transient measurements, we apply a DC ($V_{DC}$) and a chirped ($V_P$) voltages to the excitation electrode (Fig. 3a). A pulse with 14.25 MHz center-frequency and 0.2 MHz bandwidth, propages at 77 m/s group velocity. The space-time evolutions of the pulse show a stable energy transport with negligible backscattering from the two sharp corners (Fig. 3b). These results demonstrate robustness of the waveguides.

Another crucial aspect of topological insulators is the unidirectional propagation for distinct pseudospin modes. To characterize this, we fabricate another NEMM with a spin-splitter configuration consisting of four domain walls, which has been employed in several previous studies[11,12] (Fig. 4a and 4b). Such geometry allows using a simpler pseudospin selective excitation. In this configuration, the propagating direction of a pseudospin state depends on the spatial configuration of the two topological phases, $w$ = 5.5 and 6.5 µm (Fig. 4a). The pseudospin states are filtered to have a single dominant state in the input port (yellow arrow in Fig. 4a). After the signal passes the input channel, the filtered spin state mainly propagates to output port 1 and 3 (yellow arrows in Fig. 4a). We do not observe propagation to port 2 since that channel does not preserve the incident pseudospin state in the propagation direction (cyan arrow, Fig. 4a). To systematically investigate such propagation behaviors, we send voltage pulses to the excitation electrode and measure transient responses of the propagating pulses (Methods). We scan 13 sites (7 sites from the input channel and 6 sites from each output channel) near the channels' crossing point (Fig. 4b). Note that steady-state frequency response at the end of the three output ports (Fig. 4a) exhibit almost identical edge state responses due to boundary scatterings (Extended Data Fig. 5). The pulse we investigate has a 15.1 MHz center-frequency and 0.5 MHz bandwidth, which is enough to cover the broad frequency ranges of edge states. As expected, the measured signals show substantial energy transport to output 1 (Fig. 4c) and 3 (Fig. 4e), but not to output 2 (Fig. 4d), after the crossing point (white arrows in Fig. 4c to 4e). The presented results confirm that the propagation direction depends on the types of pseudospins. The use of such spin selective excitation and detection methods will enable compact, mechanical uni-directional devices.

The results we present in this work show that nanoelectromechanical metamaterials can be used as platforms for on-chip topological acoustic devices. In the future, these systems can be employed for stable ultrasound and radio frequency signal processing. With advanced nanofabrication techniques, more sophisticated structures can be realized to design other types of topological devices, for example, based on perturbative metamaterials design methods[17,18]. Moreover, frequency tunability in nanoelectromechanical resonators via electrostatic forces[32,33] can be used for electrically tunable devices[31] and actively reconfigurable topological channel[11].

**Acknowledgements**

We acknowledge partial support for this project from NSF EFRI Award No. 1741565, and the Kavli Nanoscience Institute at Caltech.

**Author contributions**

J.C. and C.D. conceived the idea of the research. J.C. designed and fabricated the samples. J.C. built the experimental setups and performed the measurements. J.C. performed the numerical simulations. J.C. and K.K performed theoretical studies. J.C., K.K. and C.D. wrote the manuscript.

**Competing Financial Interests**

Nothing to report.

**Methods**

**Sample fabrication**

The fabrication process begins with a pattern transfer by electron beam lithography and development of a PMMA resist in a MIBK:IPA=1:3 solution. The excitation electrodes, made of a Au (45 nm)/ Cr (5 nm) layer, are deposited on a 100nm-LPCVD silicon nitride ($SiN_x$)/140 nm-thermal $SiO_2$/525 μm highly-doped Si wafer, followed by a lift-off process in acetone. A second electron beam lithography step, with ZEP 520 e-beam resist, is then performed to create the pattern of etch holes (with 500 nm diameters) arranged in the extended honeycomb lattices (Fig. 1a and Extended Data Fig. 1). We use an ICP-reactive ion etch, to drill the holes on the $SiN_x$ layer. After we finish the etching of the holes, we immerse the samples in a Buffered Oxide Etchant (BOE) solution for ~ 45 - 46 minutes, to partially etch the thermal $SiO_2$ underneath the $SiN_x$ device layer. The etching duration determines the diameter of the etching circles, $r$ (Extended Data Fig.1). Detailed fabrication methods can be found in Ref. 31.

**Experiments**

The flexural motions of the membranes are measured using a home-built optical interferometer (HeNe-laser, 633 nm wavelength) with a balanced homodyne method. The measurements are performed at room temperature and a vacuum pressure $P < 10^{-6}$ mbar. The optical path length difference between the reference and the sample arms are stabilized by actuating a reference mirror. This mirror is mounted on a piezoelectric actuator which is controlled by a PID controller. The motion of the membranes is electrostatically excited by simultaneously applying DC and time-varying voltages through a bias tee (Mini-circuits, ZFBT-6GW+). The intensity of the interfered light from the reference mirror and the sample is measured using a balanced photodetector, which is connected to a high-frequency lock-in amplifier (Zurich instrument, UHFLI). The measurement position, monitored via a CMOS camera, can be controlled by moving a vacuum chamber mounted on a motorized XY linear stage.

For the dispersion curve measurements in Figure 2, we measure (at steady-state) frequency responses ranging from 10 to 20 MHz of 20 scanned sites along the edge waveguide. The scanning step is the one-dimensional lattice spacing, $a = 18$ μm. The lock-in amplifier (Zurich instrument, UHFLI) allows measuring the amplitude responses and the phase differences between the measured signal and the excitation source. To plot the frequency dispersion, we perform Fast-Fourier transformation of the amplitude × sin(phase) data. The amplitude only data and the phase-considered data are shown in Extended Data Figure 3a and 3b.

For transient measurements in Figure 3 and 4, we send a chirped signal (AWG module in UHFLI) and measure the signal with an oscilloscope (Tektronix, DPO3034). As the signal is invisible for a low excitation amplitude, we first filter the RF-output signals from the photodetector with a passive band-pass filter (6 – 22 MHz bandwidth) and average 512 data n time-domain. For robustness measurements (Fig. 3), we send a pulse containing frequency content ranging from 14 to 15 MHz, by applying $V_{DC} = 15$ V and $V_P = 75$ mV to the excitation electrode. We then perform post-signal processing to extract signals of interest, by applying a Burtterworth filter with 14.25 MHz center-frequency and 0.2 MHz bandwidth. For pseudospin-dependent transport measurements, we use a pulse (14 – 16 MHz) and applied $V_{DC} = 15$ V and $V_P = 22.5$ mV. We then apply a Burtterworth filter with 15.2 MHz center-frequency and 0.5 MHz bandwidth.

**Numerical Simulations**

We perform finite-element simulations to calculate the phononic frequency dispersion curves using COMSOL© multiphysics. We employ the pre-stressed eigenfrequency analysis module in membrane mechanics. We also consider geometric nonlinearity, to reflect the effects of residual stresses. The physical properties of $SiN_x$ used in the simulations are 3000 kg/m$^3$ density, 290 GPa Young's modulus, 0.27 Poisson ratio, and 50 MPa isotropic in-place residual stress. The lattice parameter, $a$, is chosen to be 18 µm. We calculate frequency dispersion curves for various unit cell geometries with different $w$ ranging from 5.5 µm to 6.5 µm. The center hexagon and the six corners of each unit cell are fixed, due to the presence of unetched $SiO_2$ (light grey regions in the scanning microscope images in Fig. 2b-d). The radii of the etched circles are set to $r = 4.9$ µm (Fig. 1 and Extended Data Fig. 2). We apply Bloch periodic conditions to the six sides of a unit cell, $u(\vec{r}+\vec{R}) = u(\vec{r})\exp(i\vec{q}\cdot\vec{R})$, via Floquet periodicity in COMSOL©. Here, $\vec{r}$ is a position within a unit cell, $\vec{R}$ is a lattice translation vector, and $\vec{q}$ is a wave vector. We calculate the dispersion curves along the boundary of the irreducible Brilluoin zone $\Delta$ $M_A$ $\Gamma_A$ $K_A$ (Fig. S1b)

We also numerically calculate the frequency dispersion curves of the edge states to validate the topological behaviors. As we are interested in one-dimensional dispersion along the interface, we build a strip-like super cell with 18 µm periodicity. Each topological phase ($w = 6.0 \pm 0.5$ µm) spans about ± 160 µm from the interface in the direction perpendicular to the interface. We calculate the frequency dispersion by applying one-dimensional Floquet periodic condition.

**Data availability.** The data that support the findings of this study are available from the corresponding author upon reasonable request.

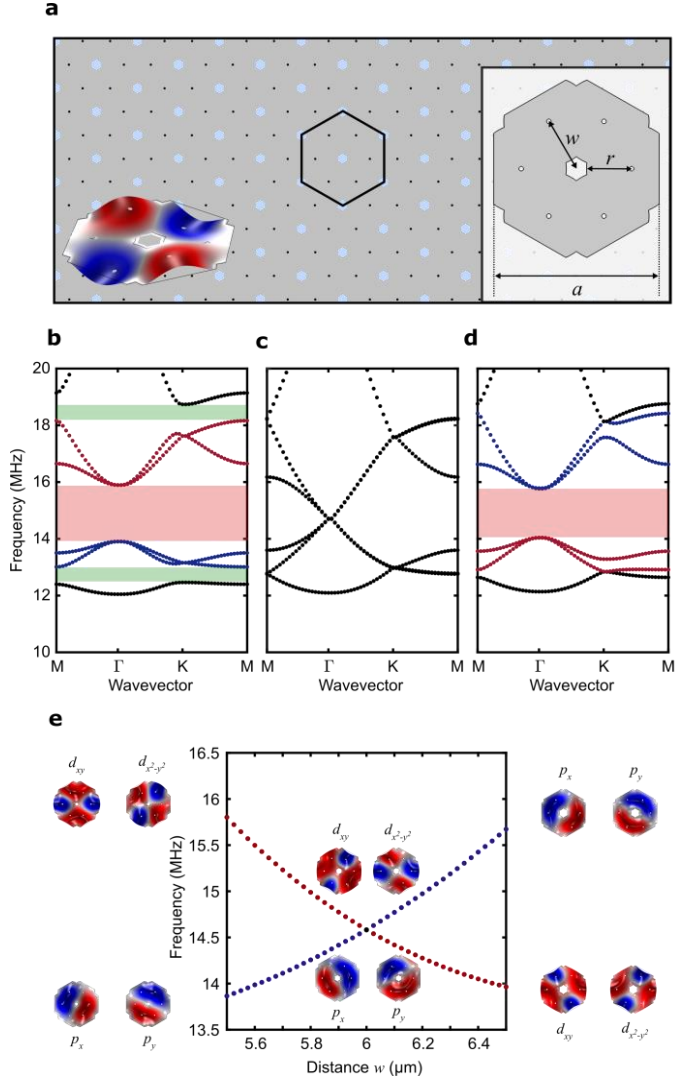

**Figure 1. Unit cell geometry and topological phase transitions.** (**a**) Schematic of a two-dimensional nanoelectromechanical metamaterial. The grey area represents the silicon nitride nanomembrane suspended over a highly doped *n*-type silicon substrate. The black dots, forming a honeycomb lattice, represent etch holes. The light blue hexagons represent the unetched thermal oxide, acting as fixed boundaries. The unit cell geometry (black solid hexagon) is shown in the right inset, with relevant parameters. An example flexural mode is shown in the left inset. The topological phases are controlled by changing $w$. (**b**), (**c**), (**d**) Frequency dispersion curves along a boundary of the irreducible Brillouin-zone MΓKM, when (**b**) $w = 5.5$ μm, (**c**), 6.0 μm, and (**d**) 6.5 μm. The red and green shaded regions correspond to topological and non-topological band gaps, respectively. (**e**) Eigenfrequencies above and below the topological band gap at the Γ point, as a function of $w$. Blue (red) dots denote the eigenfrequencies for flexural modes $p_x$ and $p_y$ ($d_{xy}$ and $d_{x^2-y^2}$). The flexural mode shapes are presented for $w = 5.5$ μm (left), 6.0 μm (middle), and 6.5 μm, respectively (right).

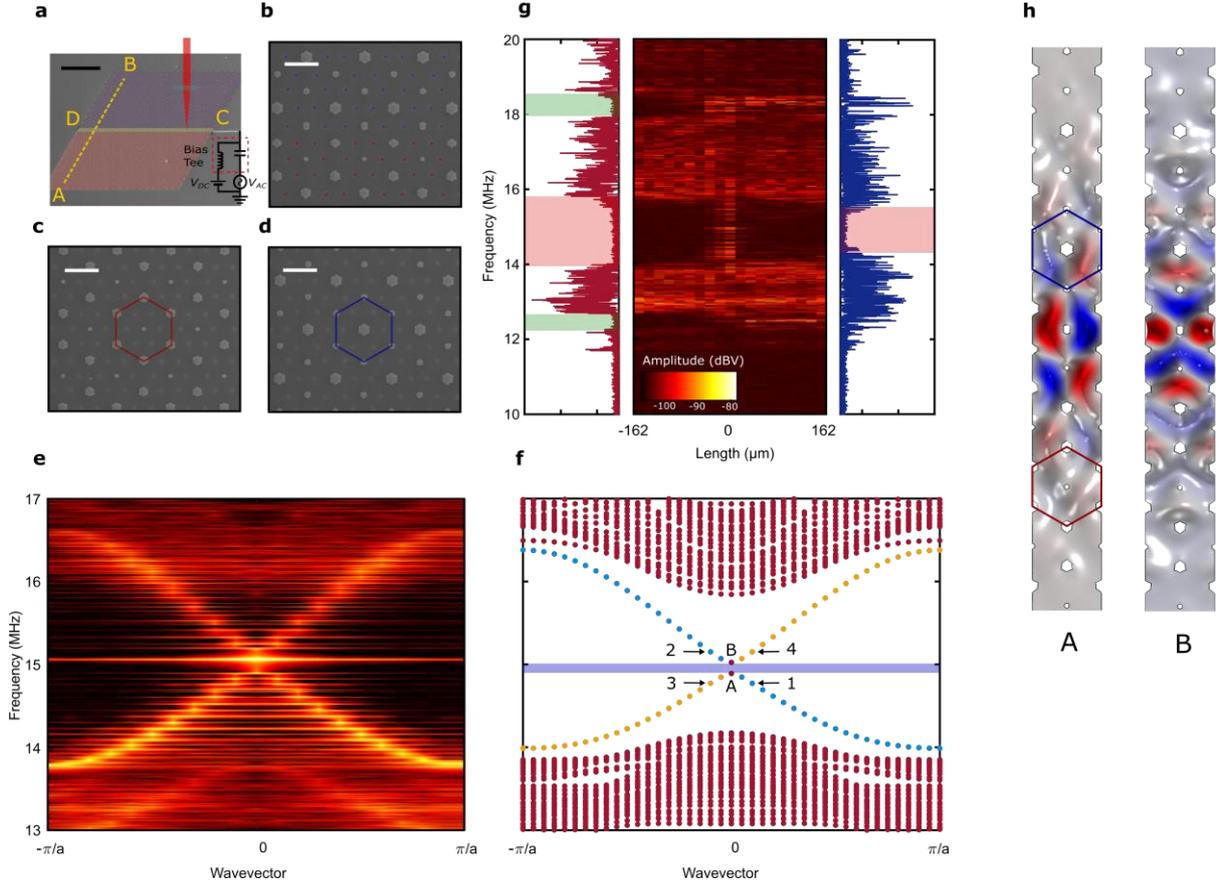

**Figure 2. Characterization of topological edge states.** (**a**) Scanning electron microscope image of a straight topological edge waveguide. The two different topological phases are shaded by blue (non-trivial) and red (trivial) false colors. Flexural membrane motions are excited by simultaneously applying DC and AC voltages ($V_{DC}$ = 2 V, $V_{AC}$ = 20 mV) to the excitation electrode via a bias tee. Scale bar, 100 μm. (**b**), (**c**), (**d**) Scanning electron microscope images of (**b**) edge region, yellow shaded strip in (**a**), (**c**) trivial lattice with $w$ = 5.5 μm, the red shaded area in (**a**), and (**c**) non-trivial lattice with $w$ = 6.5 μm, the blue shaded area in (**a**). Scale bars are 10 μm. The red and blue dots in (**b**) denote the lattice points for $w$ = 5.5 μm, and $w$ = 6.5 μm, respectively. The blue and red hexagons in (**c**) and (**d**) represent the unit cells for $w$ = 5.5 μm, and $w$ = 6.5 μm. (**e**) Experimental and (**f**) numerical frequency dispersion curves along the edge waveguide (from C to D in **a**). Yellow and light blue dots in the edge state dispersion in (**f**) represent propagating waves for up- and down- pseudospins, respectively. Time-evolution of the mode shapes at points 1,2,3, and 4 are provided in supplementary video files. (**g**) Frequency responses for 19 different sites along line A-B shown in (**a**) (middle panel). The left and right panels represent frequency responses at site A and B, respectively. The red and green shaded regions represent the band gaps. (**h**) Flexural modes for point A and B in the numerical dispersion shown in **f**. The width of the strip is 18 μm, identical to the lattice parameter $a$.

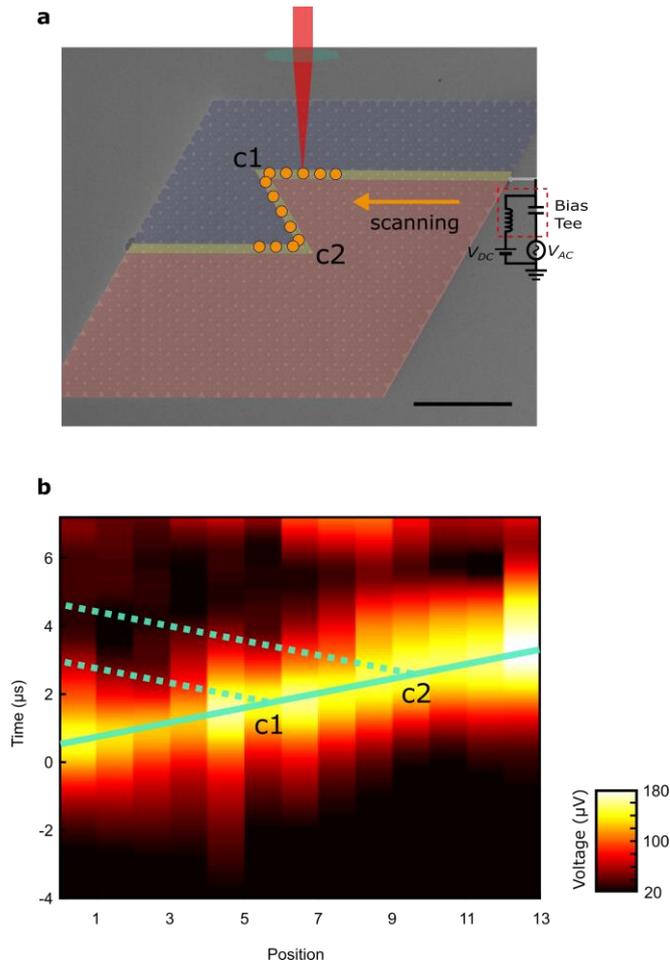

**Figure 3. Waveguide robustness against imperfections.** (**a**) Scanning electron microscope picture of a zig-zag topological edge waveguide. The red and blue shaded regions represent topologically trivial and non-trivial lattices, respectively. The flexural membrane motions are excited by simultaneously applying DC and a chirped signal with frequency content ranging from 14 MHz to 15 MHz. The applied voltages are $V_{DC} = 15$ V, $V_P = 75$ mV, here $V_P$ is the amplitude of the chirped signal. The orange dots represent measurement points. Scale bar is 100 μm. (**b**) Transient responses along the edge waveguide in a space-time domain. A pulse with 14.23 MHz center-frequency and 0.2 MHz bandwidth is considered. The position denotes the measurement points in the scanning direction shown in (**a**). c1 and c2 mark the position of the sharp corners. The solid line is added to highlight the trajectory of the propagating pulses. The two dotted lines mark the expected trajectory of the backscattered signal, if the propagating pulse were reflected from corners c1 and c2.

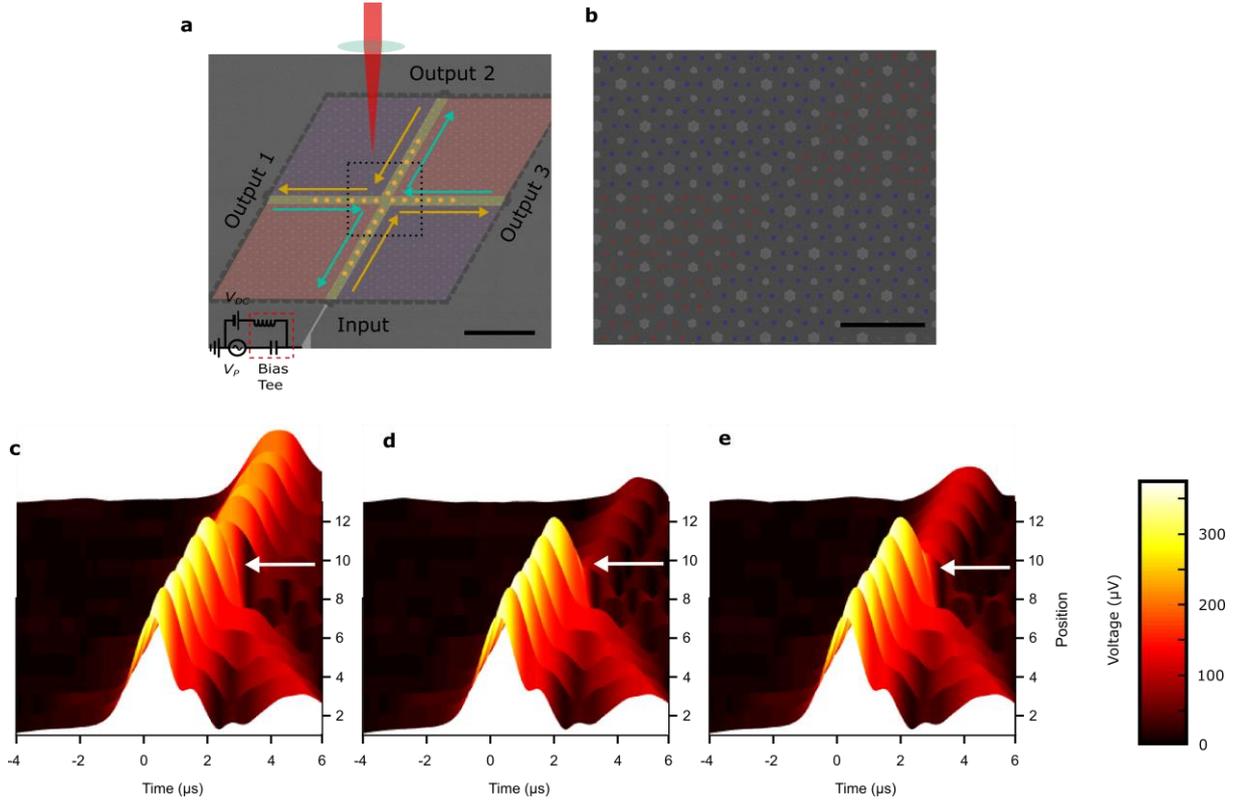

**Figure 4. Pseudospin-dependent wave propagation.** (**a**) Scanning electron microscope image of a pseudospin-filter configuration. Scale bar is 100 μm. The two different topological phases are shaded by red (trivial, $w = 5.5$ μm) and blue (non-trivial, $w = 6.5$ μm) false colors. Flexural motions are excited from the electrode ($V_{DC} = 15$ V, $V_P = 22.5$ mV). The yellow and cyan arrows represent propagating directions for different pseudospin states. (**b**) Zoomed-in region marked by the dotted square in (**a**). The red and blue dots denote the lattice points for trivial and non-trivial phases. Scale bar is 30 μm. (**c**), (**d**), and (**e**) Envelopes of propagating pulses (15.2 MHz center-frequency, 0.5 MHz bandwidth) in space-time domain. The position represents the measurement points along the edge waveguides. (**c**) Input to Output port 1, (**d**) Input to Output port 2 (**e**) Input to Output port 3. The crossing point is indicated by white arrows in (**c**), (**d**), and (**e**).